\documentclass{cimento}
\usepackage{graphicx}

\newcommand{\beq}{\begin{equation}}
\newcommand{\eeq}{\end{equation}}
\newcommand{\bea}{\begin{eqnarray}}
\newcommand{\ena}{\end{eqnarray}}

\title{On the physical processes which lie at the bases of time variability of GRBs}

\author{Remo Ruffini\from{ins:icra}\thanks{ruffini@icra.it}\ETC, 
Carlo Luciano Bianco\from{ins:icra}, 
Pascal Chardonnet\from{ins:f}\from{ins:icra}
Federico Fraschetti\from{ins:icra},
\atque 
She-Sheng Xue\from{ins:icra}}

\instlist{\inst{ins:icra} ICRA - International Center for Relativistic Astrophysic and Physics 
Department, University of Rome ``La Sapienza'' - Piazzale Aldo Moro 5, I-00185 Rome, Italy
  \inst{ins:f} Universit\'e de Savoie - LAPTH LAPP - BP110 - 74941 Annecy-le-Vieux 
Cedex, France}

\begin{document}

\maketitle

\begin{abstract}
The relative-space-time-transformation (RSTT) paradigm and the interpretation of the burst-structure (IBS) paradigm are applied to probe the origin of the time variability of GRBs. Again GRB~991216 is used as a prototypical case, thanks to the precise data from the CGRO, RXTE and Chandra satellites. It is found that with the exception of the relatively inconspicuous but scientifically very important signal originating from the initial ``proper gamma ray burst'' (P-GRB), all the other spikes and time variabilities can be explained by the interaction of the accelerated-baryonic-matter pulse with inhomogeneities in the interstellar matter. This can be demonstrated by using the RSTT paradigm as well as the IBS paradigm, to trace a typical spike observed in arrival time back to the corresponding one in the laboratory time. Using these paradigms, the identification of the physical nature of the time variablity of the GRBs can be made most convincingly. It is made explicit the dependence of a) the intensities of the afterglow, b) the spikes amplitude and c) the actual time structure on the Lorentz gamma factor of the accelerated-baryonic-matter pulse. In principle it is possible to read off from the spike structure the detailed density contrast of the interstellar medium in the host galaxy, even at very high redshift.  
\end{abstract}

It is well known that one of the most successful cognitive tools in relativistic astrophysics has been the analysis of the time structure of signals received at a variety of wavelengths. Time variabilities, however, have not always been of significance in relativistic astrophysics. In the case of pulsars, for example, only the period of the average pulsar signal and its monotonic lengthening with time have been essential in identifying pulsars as rotating neutron stars (Hewish, Bell, et al.~(1977) \cite{hgf77}). Furthermore, the modulations of the pulsar signal, periodic in time, have been essential for the identification of binary pulsars and to give the first evidence for gravitational waves (Hulse \& Taylor~(1975) \cite{ht75}). The secular variation of the shape of the pulse yields information relating the role of the spin of the neutron star and its angular momentum to additional general relativistic effects (Damour \& Ruffini~(1974) \cite{dr74}, Kramer~(2001) \cite{kr01}).  Even in this very successful example, there is a broad range of effects connected to pulsars whose role in relativistic astrophysics and fundamental physics is null. We quote, as an example, the well known time delay in the arrival time of pulsar signals, inversely proportional to the square of the radiation frequency of observation, see e.g. Rees, et al. (1975) \cite{ra75} page 26. Such an effect is {\em not} due to the mass of the photon, as one might have hoped for fundamental physics reasons (see e.g. Ohanian \& Ruffini (1994) \cite{or94}, page 117), but simply due to the dispersion by the electrons in interstellar plasma.

In approaching the analysis of GRB signals it is similarly essential to untangle information about the astrophysical system producing the GRBs, which is certainly in the realm of relativistic astrophysics, from other parts of the signal, also of similar magnitude and structure, which can instead be traced back to the environment in which the astrophysical process occurs and in that sense may very well belong to the domain of classical astronomy.

In the last decade an enormous number of papers have been written trying to link all the structures observed by the BATSE experiment on the CGRO satellite to the intrinsic properties of an unknown GRB source, whose properties should be determined, hopefully, by those observations. In the ``internal shock models'' of GRBs, which are currently very popular, it is assumed that {\em every spike} in the burst in the range $\Delta t\sim 1$ s to $\sim 50$ s is directly related to the physical properties of the ``inner engine'' (see e.g. Piran (2001) \cite{p01} and references therein). The fact that it is difficult to explain the long bursts has led the theorists working on the ``internal shock model'' to introduce a new family of models, in which the source of GRBs has a prolongued action in time. We shall see below that a simpler and very different explanation can be found.

\begin{figure}
\begin{center}
\resizebox{\hsize}{!}{\includegraphics{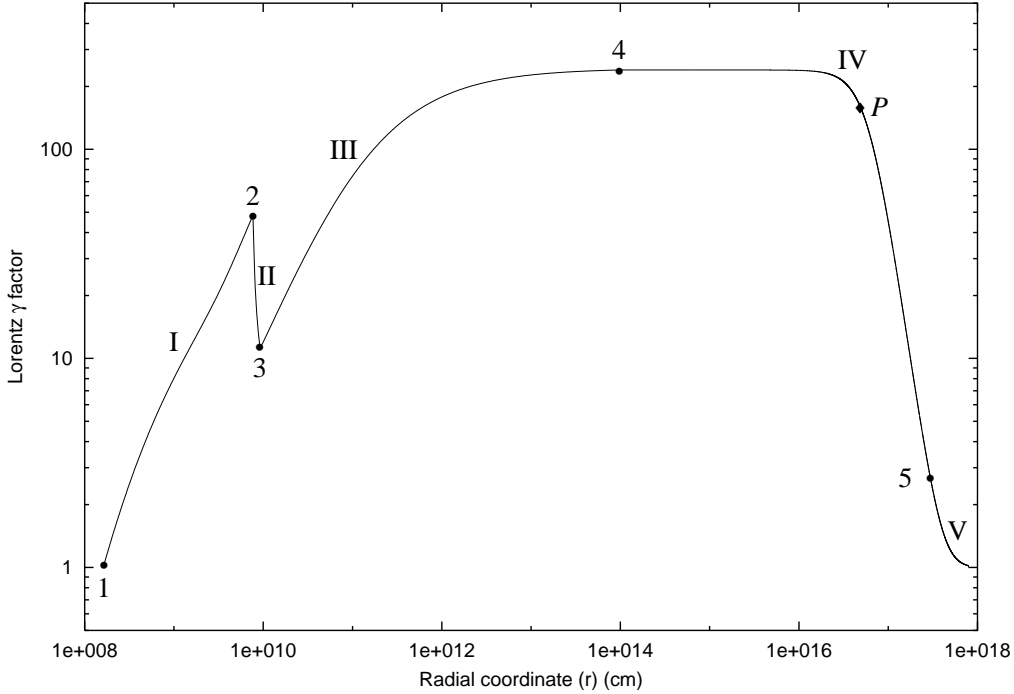}}
\caption{The Lorentz gamma factor corresponding to the different eras of GRB991216 is given as a function of the radial distance from the EMBH. Details in Letter 1.
\label{fig1}}
\end{center}
\end{figure}

The electromagnetic black hole (EMBH) model (Ruffini~(1998) \cite{rukyoto}, Preparata, et al.~(1998a) \cite{prxprl}, Preparata, et al.~(1998b) \cite{prx98}) relates the origin of the energy of GRBs to the extractable electromagnetic energy of an EMBH \cite{cr71} via the vacuum polarization process occurring during the gravitational collapse leading to the formation of an EMBH \cite{dr75}. The first step in this theory is the definition of the dyadosphere \cite{rukyoto,prxprl}, an extended region outside the EMBH horizon formed of an optically thick plasma of electron-positron pairs and radiation whose energy $E_{dya}$ is related to the mass $ \mu = M/M_\odot$ and electromagnetic parameter $ \xi = Q/\left(M\sqrt{G}\right)$ of the EMBH by the relation:\\
\begin{equation}
\displaystyle{
E_{dya} = \frac{Q^2}{2 \; r_{+}} \left(1 \; - \; \frac{r_{+}}{r_{ds}} 
\right)
\left[ 1 \; - \; \left(\frac{r_{+}}{r_{ds}}\right)^2 \right] ,
}
\end{equation} 
where  $r_{+} =  1.47 \times 10^5 \mu ( 1 \;+ \; \sqrt{1 - \xi^2} ) $ is the horizon radius and  $ r_{ds} = 1.12 \times 10^8 \sqrt{\mu \xi} $ is the dyadosphere radius and, as usual, $M$ and $Q$ are the mass-energy and charge of the EMBH and $G$ is the Newton constant of gravity. 

The evolution of this pair-electromagnetic plasma leads to the formation of a sharp pulse (the PEM pulse) that very rapidly reaches a Lorentz gamma factor of $10^2$ and higher.  The subsequent interaction of this pulse with the baryonic matter of the remnant, left over from the gravitational collapse of the protostar, and with the interstellar medium (ISM) leads to the different eras of the GRBs. It is useful to parametrize the baryonic mass $M_B$ of the remnant by introducing the dimensionless parameter $B$:
\begin{equation}
M_B c^2 = B E_{dya} .
\end{equation}

The confrontation of the theoretical model with the observational data allows an estimate for the values of the EMBH parameters. It also allows us to probe the density of the baryonic material in the remnant, in the ISM as well as in the stellar distribution within a few parsecs of the EMBH (see \cite{lett2,lett3}).

\begin{figure}
\begin{center}
\resizebox{\hsize}{!}{\includegraphics{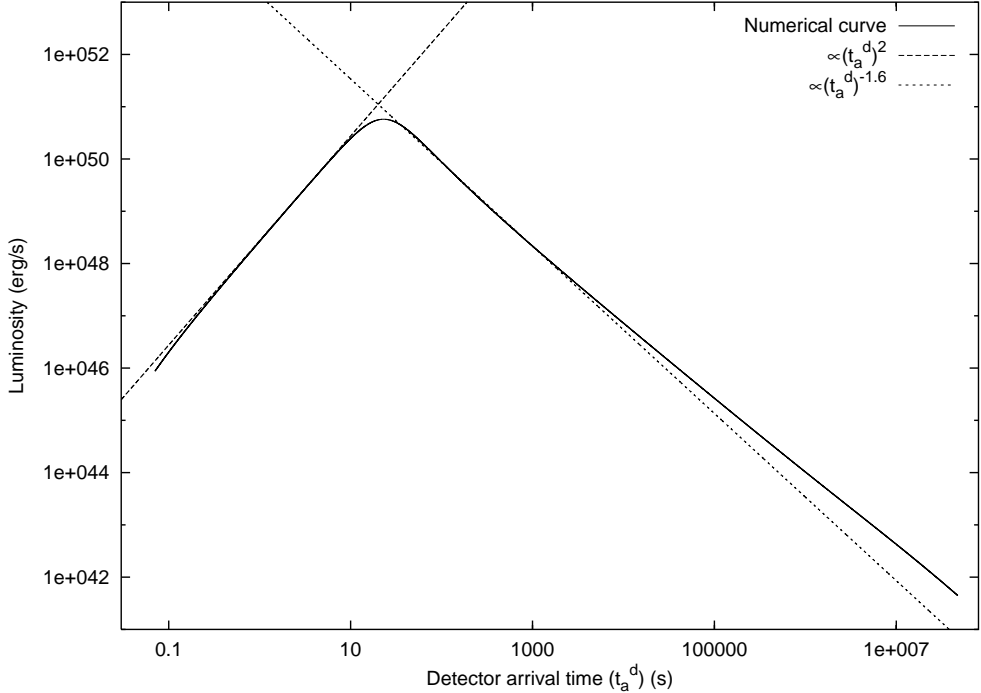}}
\caption{The flux of the afterglow of GRB991216, as computed using the best fit of the data obtained in Letter 2, is given as a function of the arrival time. The dashed (dotted) line corresponds to Eq.(\ref{F1lab}) 
(Eq.(\ref{F1labt})) in the text.
\label{fig2}}
\end{center}
\end{figure}

In Ruffini, et al.~(2001a) \cite{lett1} we presented the relative-space-time-transformation (RSTT) paradigm, leading to the
diagram relating the Lorentz gamma factor of the pulse to the space and time parametrization both in the comoving and in the laboratory frame for the case of GRB~991216 (see Fig.~\ref{fig1}). In Ruffini, et al.~(2001b) \cite{lett2} we introduced the interpretation of the burst-structure (IBS) paradigm, presenting a drastic separation between the proper-gamma-ray burst (P-GRB) and the E-APE, the ``not burst component of the GRB'', see Fig.~\ref{fig2}.

It is important to stress that the results obtained in the IBS paradigm are of general validity for a variety of GRB sources based on a single gravitational collapse event. What makes the EMBH model uniqueness testable are:
\begin{itemize}
\item the energetics \cite{rukyoto},
\item the time structure of the P-GRB \cite{rvx01},
\item the spectral information of the P-GRB \cite{bcfrx01}.
\end{itemize}

Once again we use GRB~991216 as a prototypical case due to the excellent data from the BATSE (BATSE Rapid Burst Response~(1999) \cite{brbr99}), RXTE (Corbet \& Smith~(2000) \cite{cs00}) and CHANDRA (Piro, et al.~(2000) \cite{p00}) satellites, although the conclusions will be applicable to all GRBs. In a certain sense, in this paradigm all the features of a GRB are divided in two very distinct phases:
\begin{itemize}
\item the first, prior to decoupling and ending with
emission of the P-GRB, which we shall call the ``injector 
phase'';
\item the second, the ``beam-target''
phase, in which the accelerated-baryonic-matter (ABM) pulse (the beam) interacts with the interstellar medium (the target). 
\end{itemize}

\begin{figure}
\begin{center}
\resizebox{\hsize}{!}{\includegraphics{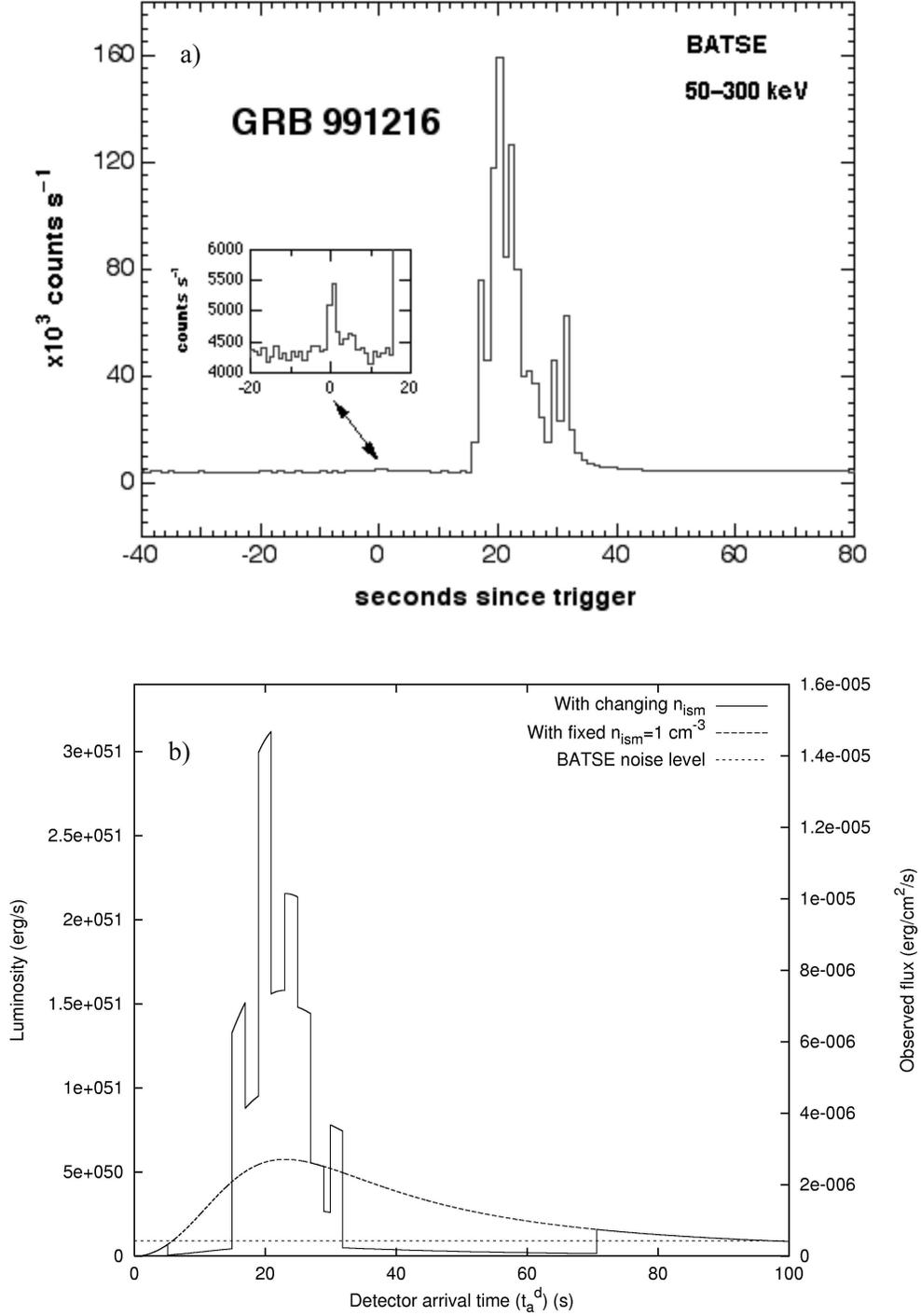}}
\caption{a) Flux of GRB 991216 observed by BATSE. The enlargement clearly shows the P-GRB (see Letter 2). 
b) Flux computed in the collision of the ABM pulse with an ISM cloud
with the density profile given in Fig.~\ref{fig4}. The dashed line indicates the emission from an uniform ISM with $ n = 1 cm^{-3} $.
The dotted line indicates the BATSE noise level.
\label{fig3}}
\end{center}
\end{figure}

The P-GRB, which is clearly identifiable in the enlargement in Fig.~\ref{fig3}a, is emitted when the condition of transparency is reached by the pair-electromagnetic-baryonic matter-pulse, the PEMB pulse. As already stressed Ruffini, et al.~(2001b) \cite{lett2}, the properties of the P-GRB are directly linked (Ruffini, et al.~(1999) \cite{rswx99}, Ruffini, et al.~(2000) \cite{rswx00}, Bianco, et al.~(2001) \cite{brx00}, Ruffini, et al. (2001g) \cite{rvx01}) to the internal properties of the GRB source and to the detailed structure and energy distribution in the dyadosphere of the EMBH (Ruffini~(1998) \cite{rukyoto}, Preparata, et al.~(1998a) \cite{prxprl}, Preparata, et al.~(1998b) \cite{prx98}). These results are essential in identifying the extractable energy  of the EMBH, introduced in Christodoulou \& Ruffini~(1971) \cite{cr71}, as the GRB energy source. The operational tool of the energy extraction process is the vacuum polarization process introduced in Damour \& Ruffini~(1975) \cite{dr75}. Similarly the intensity ratio of the P-GRB to the afterglow gives a precise measurement of the matter left in the remnant during the process of gravitational collapse of the progenitor star to the EMBH, see Ruffini, et al.~(2000) \cite{rswx00} and Ruffini, et al.~(2001b) \cite{lett2} and references therein.

All the above results clearly involve relativistic astrophysics. Let us now turn to the afterglow and apply the RSTT and IBS paradigms in order to understand its detailed time structure.

The afterglow is emitted as the ABM pulse plows through the interstellar matter engulfing new baryonic material (Ruffini, et al.~(2001f) \cite{lett6aa}). In our previous works we were interested in explaining the overall energetics of the GRB phenomena and in this sense, we have adopted the very simplified assumption that the interstellar medium is a constant density medium with $n_{ism}=1/cm^3$. Consequently, the afterglow emission obtained is very smooth in time.  We are now interested in seeing if in this frameowrk we can also explain most of the time variability observed by BATSE, all of which, except for the P-GRB, should correspond to the beam-target phase in the above paradigm.

We first recall the constitutive equations (Ruffini, et al. (1999,2000) \cite{rswx99,rswx00}, Bianco, et al. (2001) \cite{brx00}):

\begin{eqnletter}
\label{const_eq}
 \Delta E_{\rm int}  = \rho_{B_1} {V_1}\sqrt {1 + 2\gamma_1 \frac{{\Delta M_{\rm ism} c^2 }}{{\rho_{B_1} V_1 }} + \left( {\frac{{\Delta M_{\rm ism} c^2 }}{{\rho_{B_1} V_1 }}} \right)^2 }  - \nonumber \\ 
  - \rho_{B_1}{V_1} \left( 1 + \frac{{\Delta M_{\rm ism} c^2 }}{\rho_{B_1} V_1} \right) \label{heat2}\\
 \gamma_2  = \frac{{\gamma_1  + \frac{{\Delta M_{\rm ism} c^2 }}{{\rho_{B_1} V_1 }}}}{{\sqrt {1 + 2\gamma_1 \frac{{\Delta M_{\rm ism} c^2 }}{{\rho_{B_1} V_1 }} + \left( {\frac{{\Delta M_{\rm ism} c^2 }}{{\rho_{B_1} V_1 }}} \right)^2 } }} \label{dgamma2} 
\end{eqnletter}

\begin{equation}
t_a^d  = \left(1+z\right) \left(t - \frac{r}{c}\right) = \left(1+z\right) \left(t - \frac{{\int_0^t {v\left( {t'} \right)dt'}  + r_{ds} }}{c}\right)
\label{tadefz}
\end{equation}
where the quantities with the index ``$1$'' are calculated before the collision of the ABM-pulse with an elementary ISM shell of thickness $\Delta r$ and the quantities with ``$2$'' after the collision. We indicate by $\Delta E_{\rm int}$ the increase in the proper internal energy due to the collision with a single shell and by $\rho_B$ the proper energy density of the swept up baryonic matter, by $V$ the ABM pulse volume in the comoving frame, by $M_{\rm ism}$ the ISM mass swept up until radius $r$ in laboratory frame and by $\gamma$ the Lorentz factor of the expanding ABM pulse. $t_a^d$ is the arrival time of the signals on the detector, counting from the arrival of the first photon, $z$ is the cosmological redshift of the source and $t$ is the emission time of the signal, counting from the dyadosphere formation. Details are given in Ruffini, et al. (2001f) \cite{lett6aa}.

In order to proceed, we first distinguish two  different regimes in the afterglow (see Fig.~\ref{fig2}): in the first the intensity of the afterglow increases with time, in the second it decreases. The first regime goes from point 4, corresponding to the emission of the P-GRB (see Fig.~\ref{fig1}), to the point $P$, where the peak of the radiation of the afterglow is reached. During this regime, the amount of material engulfed from the interstellar medium is too small compared  to the initial kinetic energy of the ABM pulse and the Lorentz gamma factor is slowly decreasing with time, so much so that we can assume $\gamma$ is constant in this regime. The flux emitted by the afterglow is given (Ruffini, et al.~(2001f) \cite{lett6aa}), as a function of the laboratory time $t$ by

\begin{equation}
F\propto {\gamma_0}^4n_{ism}t^2 ,
\label{F1t}
\end{equation}
where $\gamma_0$ is the value of the Lorentz gamma factor at the moment of transparency. This expression can be simply expressed in terms of  the arrival time $t_a$ (Ruffini, et al.~(2001f) \cite{lett6aa}) 
\begin{equation}
F\propto {\gamma_0}^8n_{ism}\left(t_a^d\right)^2 .
\label{F1lab}
\end{equation}   

The second regime occurs as soon as the mass-energy accreted from the interstellar material is no longer negligible with respect to the initial kinetic energy of the ABM pulse (Ruffini, et al.~(2001f) \cite{lett6aa}). The flux emitted by the afterglow decreases now with the laboratory time following the law
\begin{equation}
F\propto {\gamma_P}^2{t_P}^6n_{ism}t^{-4} ,
\label{F2labt}
\end{equation}
where $\gamma_P$ is the value of the Lorentz gamma factor at the point $P$ and $t_P$ is the value of the laboratory time when the point $P$ is reached. There are actually two different peaks in the radiation flux, if the phenomenon is the spike in the laboratory frame or in the frame of an asymptotic observer comoving with the detector (see Ruffini, et al. (2001f) \cite{lett6aa} for details). Here we consider the peak in the laboratory time. The peak of the radiation occurs at a value of $\gamma$ given by (Ruffini, et al.~(2001f) \cite{lett6aa}): $\left(M_{ism}/M_B\right) \simeq 10^{-3}$ and
\begin{equation}
\gamma_P \simeq 0.67 \gamma_0 ,
\label{gammaP}
\end{equation}
where $M_B$ is the initial baryonic mass of the ABM pulse and $M_{ism}$ is the mass of the ISM engulfed by the ABM pulse at the time $t_P$. Again we can express the energy flux given in Eq.(\ref{F2labt}) as a function of the arrival time as (Ruffini, et al.~(2001f) \cite{lett6aa}) 
\begin{equation}
F\propto n_{ism}\left(t_a^d\right)^{-1.6},
\label{F1labt}
\end{equation}   
in very good agreement with the results of the BATSE, RXTE and Chandra satellites (see Letter 2).
The corresponding diagrams are summarized in Fig.~\ref{fig2}.

Once the two results presented in Fig.~\ref{fig1} and Fig.~\ref{fig2} have been understood, we can proceed to attack the specific problem of the time variability observed by BATSE.

The fundamental point is that in both regimes {\em the flux observed in the arrival time is proportional to the interstellar matter density}: any inhomogeneity in the interstellar 
medium $\Delta n_{ism}/ \overline{n}_{ism}$ will lead correspondingly to a proportional variation in the intensity  $\Delta I/ \overline{I}$ of the afterglow, which can indeed be erroneously interpreted as a burst originating in the ``inner engine''
 
\begin{figure}
\begin{center}
\resizebox{\hsize}{!}{\includegraphics{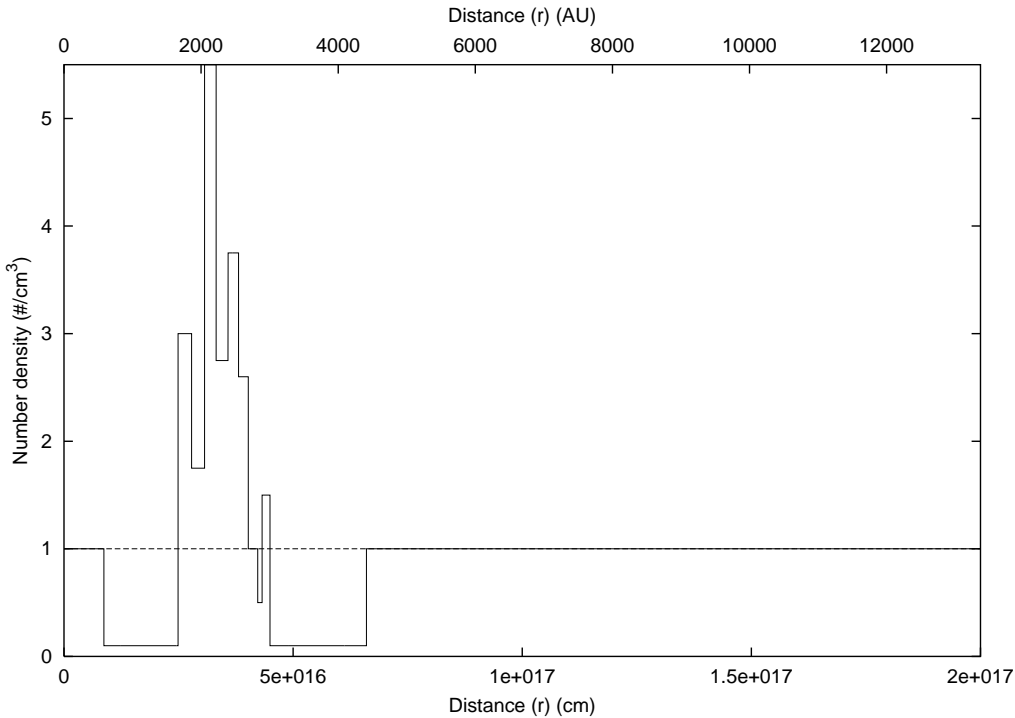}}
\caption{The density contrast of the ISM cloud profile introduced in order to fit the observation of 
the burst of GRB991216. 
The dashed line indicates the average uniform density $ n = 1 cm^{-3} $. 
\label{fig4}}
\end{center}
\end{figure}

\begin{figure}
\begin{center}
\resizebox{\hsize}{!}{\includegraphics{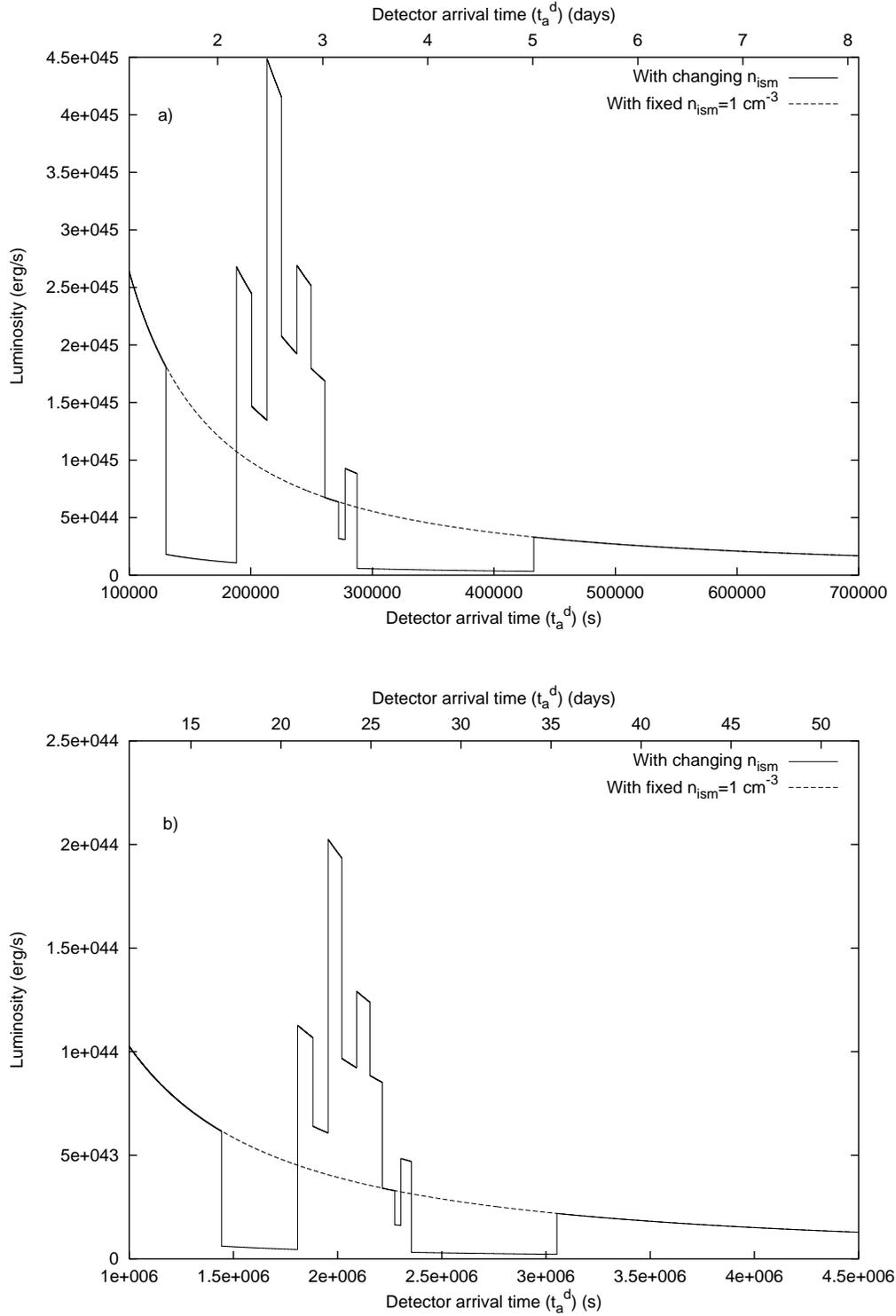}}
\caption{a) Same as Fig.~\ref{fig3}b with the ISM cloud located at a distance of $3.17\times 10^{17} cm$ from the EMBH, the time scale 
of the burst now extends to $\sim 1.58 \times 10^5 s $.
b) Same as a) with the ISM cloud at a distance of $4.71\times 10^{17} cm$ from the EMBH, 
the time scale of the burst now extends to $\sim 1.79 \times 10^6 s $.
\label{fig5}}
\end{center}
\end{figure}

There is a very significant signature of this kind of intensity contrasts: the  $\Delta I/\overline{I}$ is independent of the special moment of observation during the afterglow era and is only function of the density contrast. In particular, for the main burst observed by BATSE (see Fig.~\ref{fig3}a) we have
\beq
 \left(\Delta 
I/ \overline{I} \right) =\left(\Delta n_{ism}/n_{ism}\right) \sim 5 .
\label{disui}
\eeq        
There are still a variety of  physical circumstances which may lead to such density inhomogeneities.

The additional crucial parameter in understanding the physical nature of such inhomogeneities is 
the time scale of the burst observed by BATSE. Such a burst lasts $ \Delta t_a \simeq 20 s $ and shows substructures on a time scale of $ \sim 1s$ (see Fig.~\ref{fig3}a). In order to infer the nature of the structure emitting such a burst we must express these times scales in the laboratory time (see Letter 1). Since we are at the peak of the GRB we have $ \gamma_P \sim 159 $ (see Eq.(\ref{gammaP})) and $ \Delta t_a $ corresponds in the laboratory time to an interval
\beq
\Delta t = \gamma_P^2 \times \Delta t_a \sim 
7.5\times 10^5 s ,
\label{deltatl}
\eeq
which determines the characteristic size of the inhomogeneity creating the burst $\Delta L\sim 2.2 \times 10^{16} cm$.

It is immediately clear from Eq.(\ref{disui}) and Eq.(\ref{deltatl}) that these are the typical 
dimensions and density contrasts corresponding to a small interstellar cloud. As an explicit example we have shown in Fig.~\ref{fig4} the density contrasts and dimensions of an interstellar cloud with an {\em average density} $<n>=1/cm^3$. Such a cloud is located at a distance of $\sim 8.7\times 10^{15}cm$ from the EMBH, gives rise to a signal similar to the one observed by BATSE (see Fig.~\ref{fig3}b).

It is now interesting to see the burst that would be emitted by the interaction of the ABM pulse 
with the same ISM cloud if it were encountered at later times during the evolution of the afterglow. Fig.~\ref{fig5}a shows the structure of the burst at a distance $2.59\times 
10^{17}cm$, corresponding to an arrival time delay of $\sim 2$ days, where the Lorentz factor is now $\gamma_\star \sim 3.60$. Although the overall intensity is smaller, the intensity ratio of the burst relative to the average emission is consistent with Eq.(\ref{disui}), but the time scales of the burst are longer by a factor $ \left( \frac{\gamma_P}{\gamma_\star} \right)^2 \simeq  2 \times 10^3$. Fig.~\ref{fig5}b shows the corresponding quantities for the same ISM cloud located at a distance $3.9\times 10^{17}cm$ from the EMBH, corresponding to an arrival time delay of $\sim 1$ month, where the Lorentz gamma factor is $\sim 1.598$.

The approximations adopted in this paper in the solution of Eqs.(\ref{const_eq},\ref{tadefz}) have been explicitly presented in all details in Ruffini, et al.~(2001f) \cite{lett6aa}.

It is then clear that all the fundamental information on relativistic astrophysics about the EMBH dyadosphere as well as its formation during the process of gravitational collapse have to be inferred from the data on the propereties of the P-GRB (Bianco, et al.~(2001) \cite{brx00}, Ruffini, et al. (2001g) \cite{rvx01}).

The data on the E-APE appear to give mainly information on the structure of the ISM clouds in star-forming regions in far away galaxies.

It is then possible to carry out, very efficiently, the sort of problematic examined, within our own galaxy, by the BeppoSAX satellite (see Bocchino \& Bykov~(2000) \cite{bb00} and references therein). In these works the interstellar clouds have been examined using as ``the beam'' the  material ejected in supernova remnants, and as ``the target'' a variety of ISM clouds in our galaxy. By properly taking into account the results summarized in Fig.~\ref{fig1} and Fig.~\ref{fig2} it is in principle possible, using different GRBs, to map the interstellar matter distribution in star-forming regions in far away galaxies at arbitrary red shift.

This leads us into the domain of another science, of classical astronomy, into which the object of this work does not allow us to go today.

\end{document}